\documentclass[prd,twocolumn,superscriptaddress,preprintnumbers,balancelastpage,nofootinbib]{revtex4-1}
\usepackage{graphicx}  
\usepackage{dcolumn}   
\usepackage{bm}        
\usepackage{amssymb, amsmath}
\usepackage{slashed}   
\usepackage{lipsum}
\usepackage[usenames,dvipsnames,svgnames,table]{xcolor}
\usepackage{comment}
\usepackage[utf8]{inputenc}
\definecolor{BlueViolet}{rgb}{0.2, 0.00, 0.7}
\definecolor{Blue}{rgb}{0.15, 0.00, 0.9}
\definecolor{lightblue}{rgb}{0.15, 0.35, 0.95}
\definecolor{kitgreen}{rgb}{0,
0.58823 
, 0.50980 
}
\usepackage[
colorlinks=true, linkcolor=lightblue,citecolor=lightblue,urlcolor=kitgreen]{hyperref} 

\begin{document}
\widetext
\preprint{P3H--22--105, TTP22--064}

\title{Collider probe of heavy additional Higgs bosons solving the muon $g-2$ \\
and dark matter problems}

\author{Monika Blanke}
\email{monika.blanke@kit.edu}
\affiliation{Institute for Theoretical Particle Physics (TTP), Karlsruhe Institute of Technology (KIT),
Engesserstra{\ss}e 7, 76131 Karlsruhe, Germany}
\affiliation{ Institute for Astroparticle Physics (IAP),
Karlsruhe Institute of Technology (KIT), 
Hermann-von-Helmholtz-Platz 1, 76344 Eggenstein-Leopoldshafen, Germany}

\author{Syuhei Iguro}
\email{igurosyuhei@gmail.com}
\affiliation{Institute for Theoretical Particle Physics (TTP), Karlsruhe Institute of Technology (KIT),
Engesserstra{\ss}e 7, 76131 Karlsruhe, Germany}
\affiliation{ Institute for Astroparticle Physics (IAP),
Karlsruhe Institute of Technology (KIT), 
Hermann-von-Helmholtz-Platz 1, 76344 Eggenstein-Leopoldshafen, Germany}

\begin{abstract}
\noindent
We study the Large Hadron Collider (LHC) search potential of a $\mathbb{Z}_4$-based two Higgs doublet model which can simultaneously explain the muon $g-2$ anomaly and the observed dark matter.
The neutral scalars in the second Higgs doublet couple to $\mu$ and $\tau$ and largely contribute to the muon anomalous magnetic moment through the one-loop diagram involving $\tau$ and scalars.
An additional singlet scalar which is charged under the discrete symmetry can be a dark matter candidate.
An upper limit on the scalar mass originates from the unitarity constraint, and the $\mu\tau$ flavor-violating nature of the scalars predicts non-standard signatures at the LHC.
However, the previously proposed $\mu^\pm\mu^\pm\tau^\mp\tau^\mp$ signal via the electroweak heavy neutral scalar pair production at the LHC loses sensitivity for increasing scalar mass.
We revisit this model and investigate the LHC prospects for the single production of the $\mu\tau$ flavor-violating neutral scalar.
It is shown that the single scalar process helps to extend the LHC reach to the 1\,TeV mass regime of the scenario.
The search potential at the high energy LHC is also discussed.\\
---------------------------------------------------------------------------------------------------------------------------------
\end{abstract}
\vspace{.4cm}
\maketitle
\section{Introduction}
\label{sec:introduction}
Most  experimental results so far support the standard model (SM) of particle physics.
However, the SM falls short of explaining dark matter, the baryon asymmetry of the universe, neutrino masses and so on. 
Each of these problem has many possible solutions, and thus more experimental hints are required to specify the correct new physics (NP) scenario.
One of the most notorious and long-lived discrepancies between the SM prediction and the measurement exists in the muon anomalous magnetic moment ($a_\mu$) \cite{Aoyama:2020ynm,Muong-2:2006rrc,Muong-2:2021ojo}.
The comparison of the SM prediction with the experimental value is given as
\begin{align}
    \Delta a_\mu= a_\mu^{\rm{exp}}-a_\mu^{\rm{SM}}=(2.51\pm0.59)\times10^{-9}.
    \label{eq:mg2}
\end{align}
The SM prediction is taken from the theory white paper \cite{Aoyama:2020ynm}\footnote{See Refs.\,\cite{Aoyama:2012wk,Aoyama:2019ryr,Czarnecki:2002nt,Gnendiger:2013pva,Davier:2017zfy,Keshavarzi:2018mgv,Colangelo:2018mtw,Hoferichter:2019mqg,Davier:2019can,Keshavarzi:2019abf,Kurz:2014wya,Melnikov:2003xd,Masjuan:2017tvw,Colangelo:2017fiz,Hoferichter:2018kwz,Gerardin:2019vio,Bijnens:2019ghy,Colangelo:2019uex,Blum:2019ugy,Colangelo:2014qya}
for relevant original work.} which is mainly based on the data-driven determination of the hadronic vacuum-polarization contribution.\footnote{
We note that the estimate based on the recent lattice simulation differs and is more consistent with the measured muon $g-2$ \cite{Ce:2022kxy,Borsanyi:2020mff,Colangelo:2022vok}.
 Recent results from other lattice groups are converging towards the BMW result \cite{Ce:2022kxy,Alexandrou:2022amy}.
 However, the lattice results are in tension with the low energy $\sigma(e^+e^-\to {\rm{hadrons}})$ data \cite{Crivellin:2020zul,Keshavarzi:2020bfy,Colangelo:2020lcg}, so that further clarification is needed.
 In this paper we consider the discrepancy as quoted in Eq.\,(\ref{eq:mg2}).
} 
It is known that the discrepancy is of the same order as the electroweak contribution, i.\,e.\ a new $\mathcal{O}$(100)\,GeV  weakly coupled particle can explain the discrepancy.
However, no signal of NP at this scale has been found at the Large Hadron Collider (LHC) so far.
This fact implies that in order to explain the discrepancy in terms of NP, some enhancement mechanism in the NP contribution to $g-2$ is necessary.\footnote{See Ref.\,\cite{Crivellin:2021rbq} for a recent review.}

A popular method to enhance the $g-2$ contribution is the introduction of a new flavor-violating particle.
The dipole operator underlying $g-2$ requires a chirality flip, which corresponds to the muon mass within flavor-conserving scenarios. A one-loop contribution involving a $\mu\tau$ flavor-violating particle is instead enhanced by a factor of $m_\tau/m_\mu\simeq17$ \cite{Nie:1998dg,Diaz:2000rsq,Baek:2001kca,Iltan:2001nk,Wu:2001vq,Assamagan:2002kf,Omura:2015nja,Omura:2015xcg,Heeck:2016xkh,Altmannshofer:2016brv,Chiang:2017vcl,Dev:2017ftk,Iguro:2018qzf,Abe:2019bkf,Wang:2019ngf,Iguro:2019sly,Crivellin:2019dun,Bauer:2019gfk,Cornella:2019uxs,Iguro:2020qbc,Iguro:2020rby,Jana:2020pxx,Wang:2021fkn,Buras:2021btx,Ghosh:2021jeg,Hou:2021sfl,Han:2022juu,Kriewald:2022erk,Asai:2022uix}.\footnote{Due to the loop function, scalar mediators receive a further enhancement.}
This mechanism can lift the mass scale of the new particle by more than a factor of four.
However, lepton flavor-violating (LFV) interactions are stringently constrained and easily spoil the model if the particle also has lepton flavor-conserving couplings.
Therefore one needs to ensure the absence of flavor-diagonal couplings for the $\tau$ mass enhanced muon $g-2$ solution to be viable.

This specific coupling alignment can be realized by a discrete $\mathbb{Z}_4$ flavor symmetry within the two Higgs doublet model (2HDM) \cite{Abe:2019bkf}.\footnote{Note that a $\mathbb{Z}_4$-symmetric 2HDM always carries an accidental $U(1)$ symmetry \cite{Ferreira:2010ir}, however further extensions of the scalar sector can break the latter symmetry \cite{Ivanov:2013bka}.}
In this model the $g-2$ contribution is proportional to the $\mu\tau$ LFV coupling and the mass difference of the additional neutral scalars.
Recently it was proposed that the singlet scalar extension of the model can explain the relic density of the dark matter (DM) through the thermal freeze-out mechanism \cite{Asai:2022uix}.
The $\mathbb{Z}_4$ symmetry is then used both to stabilize the DM candidate and also to realize the flavor alignment.

\begin{figure*}[t]
\begin{center}
 \includegraphics[width=0.22\textwidth]{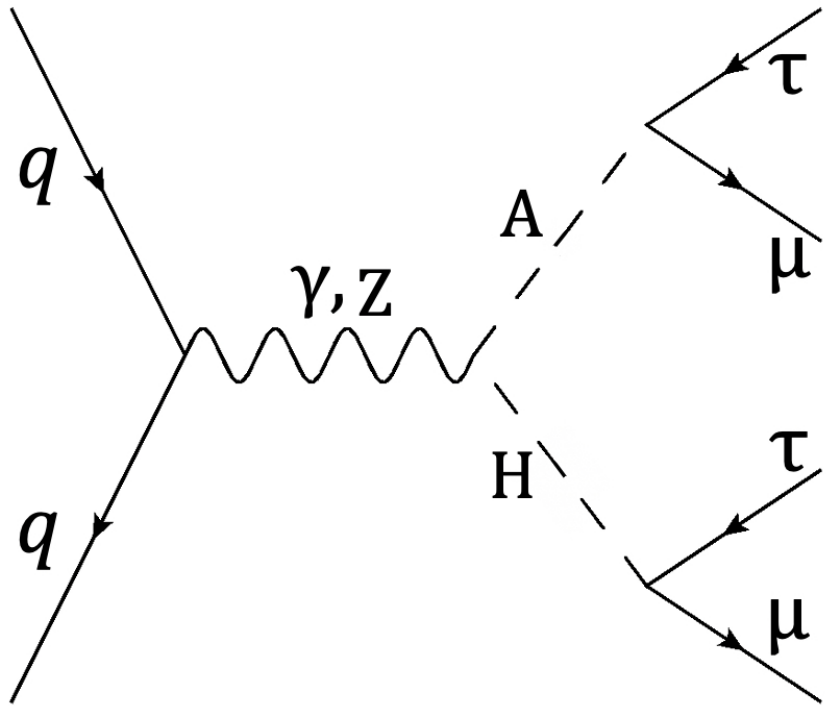}
~~~~~%
 \includegraphics[width=0.25\textwidth]{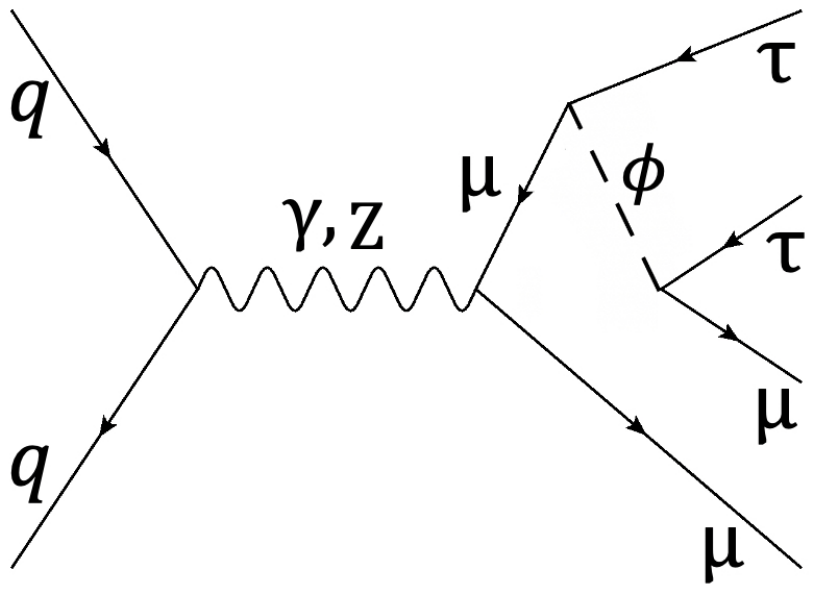}
~~~~~%
 \includegraphics[width=0.24\textwidth]{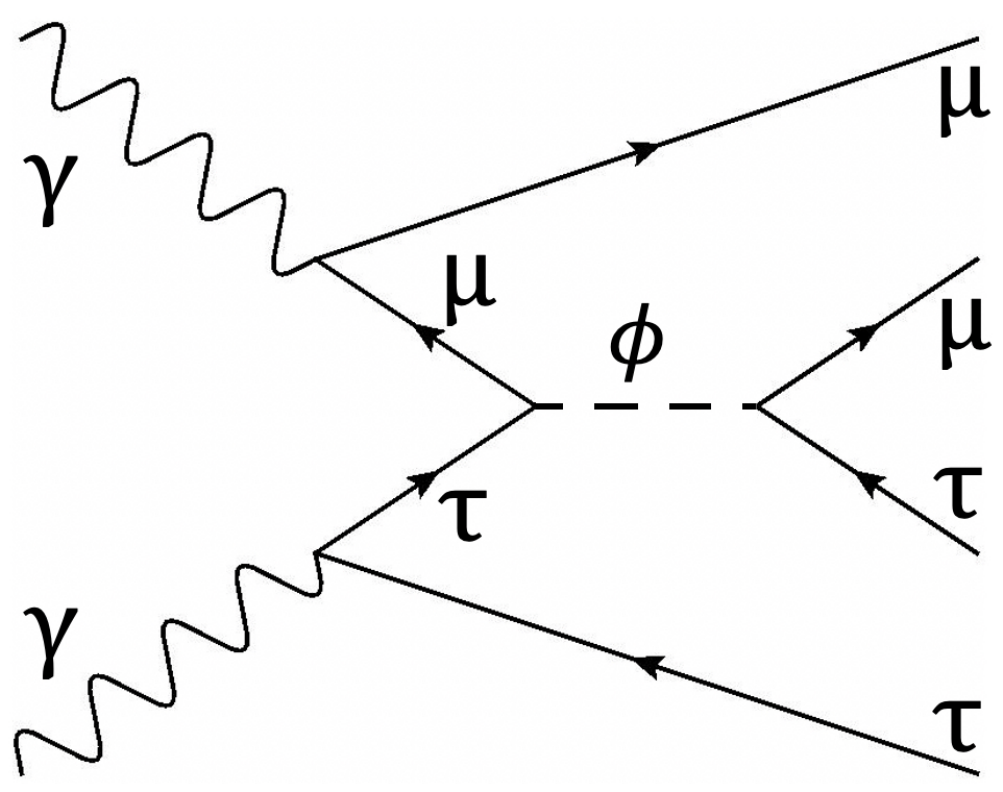}
 \vspace{-.2cm}
\caption{Representative Feynman diagrams that contribute to the $\mu^\pm\mu^\pm\tau^\mp\tau^\mp$ signal at the LHC.
The left diagram displays the electroweak pair production channel.
The middle and right diagrams correspond to the single production process where $\phi$ denotes $A$ or $H$.
In addition, there are also contributions obtained by exchanging $\mu$ and $\tau$ which are included in our numerical calculation.
  }
    \label{fig:dia}
    \vspace{-.5cm}
\end{center}
\end{figure*}

Since the new scalars are quark-phobic within the $\mathbb{Z}_4$-based model, their production cross section at the LHC is not large.
However, the unique coupling structure predicts that the neutral scalars $\phi = A,H$ decay into $\mu^\pm\tau^\mp$.
Previously we pointed out the smoking-gun signature of a $\mu^\pm\mu^\pm\tau^\mp\tau^\mp$ final state via electroweak scalar pair production (left of Fig.\,\ref{fig:dia}) with a special focus on the case where all Yukawa and scalar potential couplings are smaller than one \cite{Iguro:2019sly}.
We argued that the full Run 2 data set can test the model up to 500\,GeV scalar mass thanks to the very unique double $\mu\tau$ LFV resonance nature of the signal events.
However, if we accept relatively large coupling of $\mathcal{O}$(1), the model can still explain the discrepancy with 1\,TeV scalars.

In this paper we revisit the model's collider prospects in the presence of larger couplings.
The pair production cross section is governed only by the electroweak coupling and decreases rapidly when the scalars get heavier.
We thus propose the single scalar production process (middle and right of Fig.\,\ref{fig:dia}) to assist to cover the heavier scalar scenario.
To search for the heavy lepto-philic bosons, it is known that the inclusion of photon-initiated processes is important \cite{Iguro:2020qbc}.
We combine those processes and evaluate the search potential at the future LHC.

The layout of the paper is given as follows. In Sec.\,\ref{sec:model}, we briefly introduce our setup of the 2HDM and review the muon $g-2$ explanation.
There we determine how heavy the scalar can be and discuss relevant constraints.
In Sec.\,\ref{sec:collider}, we focus on  the collider phenomenology and show the impact of the single scalar production process to evaluate the future LHC reach.
Sec.\,\ref{sec:summary} is devoted to the summary and discussion.

\section{Model, muon $g-2$ and dark matter}
\label{sec:model}
We consider a two Higgs doublet model with an additional scalar singlet ($S$) and a discrete $\mathbb{Z}_4$ symmetry under which the Higgs and lepton fields transform as given in Tab.\,\ref{tab:model}.
The gauge charge assignments of other SM fields, e.\,g.\ quarks, are the same as in the SM, and they trivially transform under $\mathbb{Z}_4$.\footnote{In order to obtain realistic neutrino masses and mixings the model needs to be extended. See Ref.\,\cite{Asai:2022uix} for details. 
}

\begin{table}
\begin{center}
\scalebox{0.93}{
  \begin{tabular}{c|c|c|c|c|c} \hline
    Field &$H_1$&$H_2$&$(L_e,\,L_\mu,\,L_\tau)$&$(e_R,\,\mu_R,\,\tau_R)$ &$S$  \\ \hline 
   SM gauge &$(1,2)_{1/2}$&$(1,2)_{1/2}$&$(1,2)_{-1/2}$&$(1,1)_{-1}$&$(1,1)_0$ \\ \hline
   $\mathbb{Z}_4$ &$1$&$-1$&$(1,\,i,\,-i)$&$(1,\,i,\,-i)$&$i$ \\ \hline
  \end{tabular}
  }
  \caption{Relevant field content and charge assignment of the model.
  The notation of SM gauge quantum numbers is given as $(SU(3)_C,~SU(2)_L)_{U(1)_Y}$.
 }
  \label{tab:model}
\end{center}   
\vspace{-.45cm}
\end{table}

We assume the $\mathbb{Z}_4$ symmetry to be unaffected by electroweak symmetry breaking, so that the two Higgs doublets $H_{1,2}$ are in the Higgs basis~\cite{Georgi:1978ri,Donoghue:1978cj} in which only one Higgs doublet has a non-vanishing vacuum expectation value (vev) of $v \simeq 246$\,GeV.
In this basis, the two Higgs doublets can be decomposed as
\begin{eqnarray}
  H_1 =\left(
  \begin{array}{c}
    G^+\\
    \frac{v+h+iG}{\sqrt{2}}
  \end{array}
  \right),~~~
  H_2=\left(
  \begin{array}{c}
    H^+\\
    \frac{H+iA}{\sqrt{2}}
  \end{array}
  \right),
\label{HiggsBasis}
\end{eqnarray}
where $G^+$ and $G$ are the SM Nambu-Goldstone bosons, and $H^+$ and $h$ are a charged Higgs boson and the discovered CP-even Higgs boson, respectively.
$H$ and $A$ correspond to additional neutral scalars. 
Note that a non-zero vev of the singlet $S$ would spontaneously break the $\mathbb{Z}_4$ symmetry. The presence of $\langle S\rangle \ne 0$ would not alter the phenomenology discussed in the present paper, hence we do not discuss this possibility further.

The scalar potential of our model is given by \cite{Asai:2022uix}
\begin{align}
  V&=M_{1}^2 H_1^\dagger H_1+M_{2}^2 H_2^\dagger H_2+\lambda_1(H_1^\dagger H_1)^2 + \lambda_2(H_2^\dagger H_2)^2 \nonumber \\
&+\lambda_3(H_1^\dagger H_1)(H_2^\dagger H_2)
+\lambda_4 (H_1^\dagger H_2)(H_2^\dagger H_1)
\nonumber \\
&+\frac{\lambda_5}{2}\left[(H_1^\dagger H_2)^2 +{\rm h.c.}\right]\nonumber \\
&+M_S^2 |S|^2
+\lambda_S |S|^4
+\left[\lambda'_S S^4 +{\rm h.c.}\right] 
+\lambda_{S1}(H_1^\dagger H_1) |S|^2 \nonumber \\
&+\lambda_{S2}(H_2^\dagger H_2) |S|^2
+\kappa \left[(H_1^\dagger H_2)S^2 +{\rm h.c.}\right]\,.\label{eq:potential}
\end{align}
Since the mass spectrum of the scalars is of crucial importance for the muon $g-2$ as well as the collider phenomenology, we explicitly show the mass relations:
\begin{align}
  m_h^2& = \lambda_1 v^2,~~~
  m_A^2 = M_{2}^2+\frac{\lambda_3+\lambda_4-\lambda_5}{2} v^2,\nonumber \\
  m_H^2& =m_A^2+\lambda_5 v^2,~~~
  m_{H^\pm}^2 = m_A^2-\frac{\lambda_4-\lambda_5}{2} v^2,
  \nonumber \\
  m_S^2 & = M_S^2 + \frac{\lambda_{S1}}{2}v^2.
  \label{eq:Higgs_spectrum}
\end{align}
Note that due to the $\mathbb{Z}_4$ charge assignments, neither $h$ nor $S$ mix with the neutral components $H,A$ of $H_2$.
For later convenience we define the mass difference of the heavy neutral scalars $H,A$ as
\begin{align}
\Delta_{H-A}& = m_H-m_A\nonumber \\
&\simeq 50\,{\rm{GeV}}\left(\frac{\lambda_5 }{1.5}\right)\left(\frac{1800\,{\rm{GeV}}}{m_H+m_A}\right).
\label{Eq:mass_l5}
\end{align}
Note that $\lambda_5\ge0$ corresponds to $m_H\ge m_A$. 
The mass difference $\Delta_{H-A}$ decreases for heavier scalars, since it is proportional to the $SU(2)_L$ breaking.

\begin{figure*}[t]
 \includegraphics[width=0.45\textwidth]{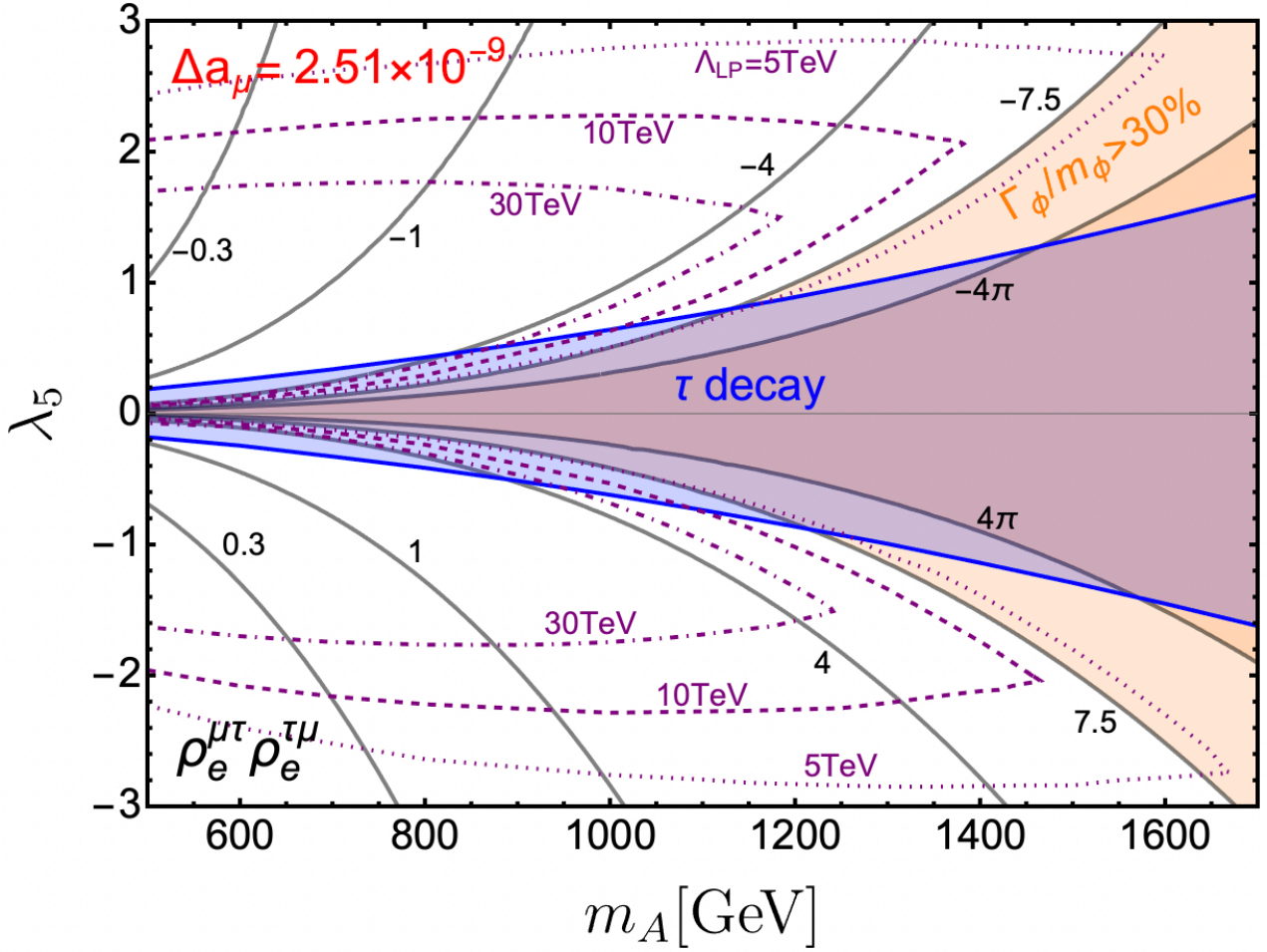}
     \vspace{-.3cm}
\caption{The black contours show the value of $\rho_e^{\mu\tau}\rho_e^{\tau\mu}$ to explain the central value of $\alpha_\mu$ in the $m_A$ vs. $\lambda_5$ plane.
The blue region is excluded by the lepton flavor universality of $\tau$ decays.
The orange region corresponds to $\Gamma_\phi/m_\phi\ge30\%$. The purple contours depict the cutoff scale of the model.    
}
    \label{fig:param}
    \vspace{-.4cm}
\end{figure*}

Thanks to the $\mathbb{Z}_4$ symmetry the Yukawa sector of the model contains only two new parameters, significantly less than in the general 2HDM. 
Following the notation in Ref.\,\cite{Omura:2015nja}, the Yukawa sector of the model based on the $\mathbb{Z}_4$ charge assignment, in addition to the SM part, is given as
\begin{align}
    -{\mathcal{L}}_Y= \rho_e^{\mu\tau} \overline{L}_\mu H_2 \tau_R+\rho_e^{\tau\mu} \overline{L}_\tau H_2 \mu_R+{\rm{h.c.}},
\end{align}
where $\rho_e^{\mu\tau}$ and $\rho_e^{\tau\mu}$ are free parameters. The scalar singlet $S$ does not couple to the SM fermions.

In this model, a sizable contribution to $\Delta a_\mu$ is generated 
via the one-loop diagram mediated by the extra neutral Higgs bosons $H$ and $A$.
The $\tau$ mass enhanced contribution\footnote{We also include non-$m_\tau$ enhanced terms of the neutral scalar loop diagram numerically, however their impact is small in our case. 
The $H^\pm$-loop contribution does not have an $m_\tau$ enhancement and thus its numerical impact is also small.} is given as \cite{Omura:2015nja,Omura:2015xcg}
\begin{align}
\Delta a_\mu
&\simeq\frac{m_\mu m_\tau\rho_{e}^{\mu\tau}\rho_{e}^{\tau\mu}}{16\pi^2}\left( \frac{\ln{\frac{m_H^2}{m_\tau^2}}-\frac{3}{2}}{m_H^2}-\frac{\ln{\frac{m_A^2}{m_\tau^2}}-\frac{3}{2}}{m_A^2}\right)\nonumber\\
&\simeq -2.5\times10^{-9}\left(\frac{\rho_{e}^{\mu\tau}\rho_{e}^{\tau\mu}}{1.0}\right)\left(\frac{\lambda_5}{1.0}\right) \left(\frac{700\,\rm{GeV}}{m_A}\right)^4,
\label{eq:gm2}
\end{align}
where Eq.\,(\ref{Eq:mass_l5}) is used to derive the second relation. Notice the dependence on the product of the two new Yukawa couplings, $\Delta a_\mu \propto \rho_e^{\mu\tau}\rho_e^{\tau\mu}$.
For given masses $m_{H,A}$, the NP effect in $a_\mu$ is thus largest if both couplings $\rho_e^{\mu\tau}$ and $\rho_e^{\tau\mu}$ are large. As we are interested in the heaviest possible  scenario, we thus set $|\rho_e^{\mu\tau}|=|\rho_e^{\tau\mu}|$ in the following.

Fig.\,\ref{fig:param} shows the value of $\rho_e^{\mu\tau}\rho_e^{\tau\mu}$ required to explain the central value of the discrepancy with black contours.
If we allow for large Yukawa couplings, heavy scalars of $\mathcal{O}$(1)\,TeV can explain the muon $g-2$ discrepancy. 
Furthermore the product $\rho_{e}^{\mu\tau}\rho_{e}^{\tau\mu}\lambda_5$ must be negative to obtain a positive contribution to $\Delta a_\mu$.
In summary we find that the parameters relevant for $\Delta a_\mu$ are $\rho_{e}^{\mu\tau}\rho_{e}^{\tau\mu}$, $m_A$ and $\lambda_5$.
It is noted that the $\tau$ mass enhanced $g-2$ contribution picks up the $SU(2)_L$-breaking effect and is proportional to $m_A^{-4}$.

The charged Higgs mass is set to $m_{H^\pm}=m_A$ $(m_{H^\pm}=m_H)$ for $\lambda_5\ge0$ ($\lambda_5\le0$) to respect the constraints from electroweak oblique parameters \cite{ParticleDataGroup:2020ssz} and vacuum stability \cite{Sher:1988mj,Muhlleitner:2016mzt}.
The size of the couplings is bounded from above by the requirement that the theory remains perturbative, we hence choose $|\lambda_i|\le 4\pi$ and $|\rho_e^{\mu\tau}|\le \sqrt{4\pi}$~\cite{Abe:2019bkf}.
Furthermore, the perturbative unitarity condition for which we require tree-level unitarity of 2$\,\to\,$2 processes is imposed \cite{Lee:1977eg,Kanemura:1993hm,Arhrib:2000is}.
Even if the couplings satisfy those constraints at the mass scale of the additional scalars, renormalization group (RG) running effects alter them and the theory would become non-perturbative at a high energy scale with $\mathcal{O}(1)$ couplings.
In order to keep the model under perturbative control, additional particles and their interactions at the scale $\Lambda_\text{LP}$ would be necessary. Such $\mathcal{O}(10\,\text{TeV})$ new particles can in principle contribute to the muon $g-2$, however the effect is small since it scales as $1/m_{\Lambda_\text{LP}}^2$. Similarly, they are out of reach of direct LHC searches. 
Finally, the contribution of the new particles to the beta functions and their impact on the RGE evolution highly depends on the nature of the new particles, e.g. their coupling structure and spin. 
To quantify this, we solve the coupled RG equations (RGEs) for the $\beta$ functions of the SM third generation fermion Yukawa couplings, $\lambda_i$,  $\rho_e^{\mu\tau}$ and $\rho_e^{\tau\mu}$, and then determine the cut-off scale, $\Lambda_{\rm{LP}}$, where LP stands for Landau pole.\footnote{See Refs.\, \cite{Abe:2019bkf,Iguro:2022tmr} for the $\beta$ functions. At the initial scale, we set $\lambda_2,~\lambda_3\ll1$ to maximize the cut-off scale.
It is noted that there are typos in the $\beta$ functions of Ref.\,\cite{Abe:2019bkf}.
}
These conditions imply the existence of an upper mass limit for the heavy scalars.
In Fig.\,\ref{fig:param} dash-dotted, dashed and dotted lines in purple depict $\Lambda_{\rm{LP}}=30\,$TeV, 10\,TeV and 5\,TeV.
It is observed that if we require the theory to be perturbative up to 30\,(5)\,TeV, the upper limit on the scalar mass is given as 
\begin{align}
m_A\le1250~(1650)\,{\rm{GeV}}.
\label{eq:thUL}
\end{align}
The large $\lambda_5$ case is constrained by the 2$\,\to\,$2 unitarity bound while the small $\lambda_5$ region is disfavored by the unitarity constraint on the RGE-evolved Yukawa couplings.

The presence of a charged Higgs with a sizable product of the relevant Yukawa couplings can modify the decay rate of $\tau\to\mu\nu\overline\nu$.
Following Ref.\,\cite{Iguro:2018qzf}, lepton flavor universality in $\tau$ decays puts an upper limit on the interaction:
\begin{align}
    \left| \frac{\rho_e^{\mu\tau}\rho_e^{\tau\mu}}{1.9}\right|\left( \frac{700\,{\rm{GeV}}}{m_{H^\pm}}\right)^2\le 1.
\end{align}
The corresponding exclusion is shown in blue in Fig.\,\ref{fig:param}.\footnote{The Belle II experiment will improve the sensitivity, however, a quantitative evaluation is not available \cite{Belle-II:2018jsg}.}
Furthermore the one-loop corrections to $Z$ and Higgs boson couplings to $\tau\overline{\tau}$ and $\mu\overline{\mu}$ are known to be less constraining.  

Since we are interested in the collider sensitivity utilizing resonant production of the new scalars $\phi = A,H$, their width-to-mass ratio is important.
This ratio is approximately given as $\Gamma_\phi/m_\phi\sim |\rho_e^{\mu\tau}|^2 \times 4\%$. 
We assume that the narrow width approximation is valid up to $30\,\%$.
The parameter region which predicts $\Gamma_\phi/m_\phi \ge 30\,\%$ is shown in orange  in Fig.\,\ref{fig:param}. Depending on the desired accuracy for LHC cross-section predictions, a more conservative limit on the validity of the narrow width approximation may be in order. 
However, since the  cross section is a steeply falling function of $m_A$, as seen in the left panels of Figs.\ \ref{fig:14} and \ref{fig:27}, and we are interested in the collider reach in terms of the mass $m_A$, even a somewhat imprecise prediction of the cross section using the narrow width approximation will yield sufficiently precise results for the (HL/HE-)LHC reach.

As shown in Ref.\ \cite{Asai:2022uix} and mentioned in the introduction, the singlet scalar $S$ is a viable dark matter candidate. In the early universe $S$ is in thermal equilibrium with the SM fields through its interaction with the scalar doublets, see eq.\ \eqref{eq:potential}.
Note that in Ref.\,\cite{Asai:2022uix} the coupling of $S$ to the SM Higgs doublet, $\lambda_{S1}$, has been set to zero and $S$ has no vacuum expectation value. $S$ then couples only to the new heavy scalars $H^\pm, H, A$ which serve as $\mu\tau$-flavoured mediators. It is then possible to explain the relic abundance with the thermal freeze-out mechanism.
Thanks to the $\mathbb{Z}_4$ charge assignment, the singlet scalar does not couple to the nucleon at the tree level, evading the stringent constraints from direct detection experiments. Similarly, due to the absence of couplings to SM particles, $S$ is not produced directly at colliders.
However, one-loop $Z$ penguin induced DM-nucleon scattering can constrain the model. The viable mass range for $S$ is broad and the next generation experiments are important to probe the interesting parameter space. More details on the phenomenology of $S$ and its dark matter interpretation can be found in \cite{Asai:2022uix}. 

We stress that while the explanation of the $g-2$ anomaly in terms of $A,H$ contributions is independent of the introduction of $S$ and its dark matter interpretation, the reverse statement is not true. On the one hand, dark matter stability is guaranteed by the same $\mathbb{Z}_4$ symmetry that is responsible for the lepton flavour-violating coupling structure of the heavy scalars. On the other hand, since the heavy scalars serve as mediators between the visible and dark sectors of the model, their mass and interaction strength with the SM leptons directly influences the dark matter phenomenology of the model.

In passing we note that charged scalar pair production is probed by the left-handed slepton search \cite{ATLAS:2019lng,ATLAS:2019lff}.
However, it is difficult to constrain O(1)\,TeV $H^\pm$ with BR$(H^\pm\to\mu\nu)\simeq0.5$ even at the HL-LHC.

\section{LHC probe of TeV scalar scenario}
\label{sec:collider}

In this section we investigate the LHC sensitivity to our model.
Ref.\,\cite{Iguro:2019sly} pointed out the smoking-gun signature  $\mu^\pm \mu^\pm \tau^\mp\tau^\mp$ via pair production of the heavy neutral scalars $HA$, each decaying into $\mu^\pm\tau^\mp$. This signature violates lepton flavour and is therefore free from irreducible SM backgrounds. Note that while  the final state $\mu^\pm \mu^\mp \tau^\pm\tau^\mp$ is produced at comparable rates, the latter is lepton flavour-conserving and therefore subject to sizeable SM background from, e.\,g., $ZZ$ production. The signature $\mu^\pm \mu^\mp \tau^\pm\tau^\mp$ is thus less suitable as a discovery channel---however in case of a $\mu^\pm \mu^\pm \tau^\mp\tau^\mp$ discovery it could potentially be used to validate the underlying model. In the present paper we focus on the discovery prospects of the lepton flavour-violating signature $\mu^\pm \mu^\pm \tau^\mp\tau^\mp$.

As a first step we  extend the previous study \cite{Iguro:2019sly}  and evaluate the high luminosity (HL)-LHC reach of the model based on the electroweak pair production of the scalars.
Ref.\,\cite{Iguro:2019sly} focused on the weakly-coupled scenario with $|\lambda_5|,~|\rho_{e}^{\mu\tau}|,~|\rho_{e}^{\tau\mu}|\le1$, and thus $m_A\le700\,$GeV was considered.
It was argued that the very distinctive $\mu^\pm\mu^\pm\tau^\mp\tau^\mp$ final state via decays of an electroweakly produced pair of neutral scalars is useful to test the model and it was shown that $139\,$fb$^{-1}$ of the data can probe the scenario with $m_A\lesssim 500\,$GeV.
The electroweak production cross section depends only on the heavy scalar masses.

\begin{figure*}[t]

 \includegraphics[width=0.51\textwidth]{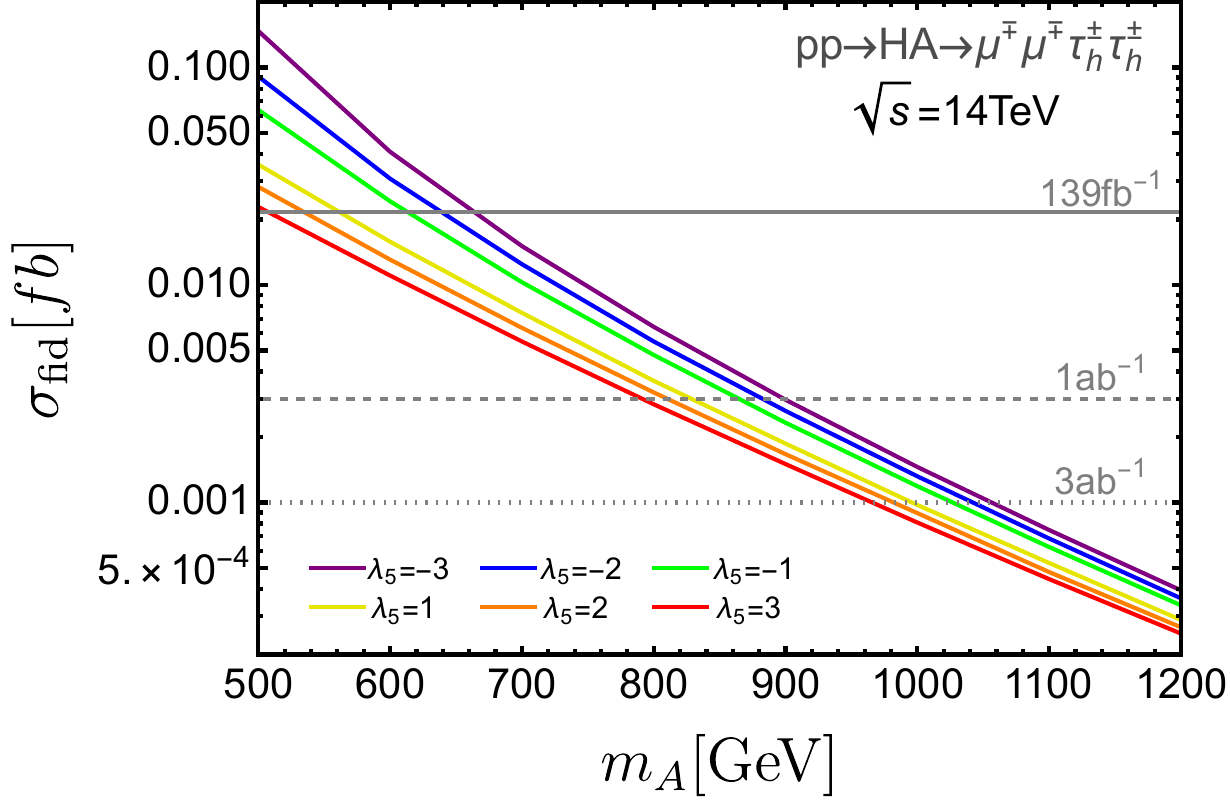}~~
  \includegraphics[width=0.47\textwidth]{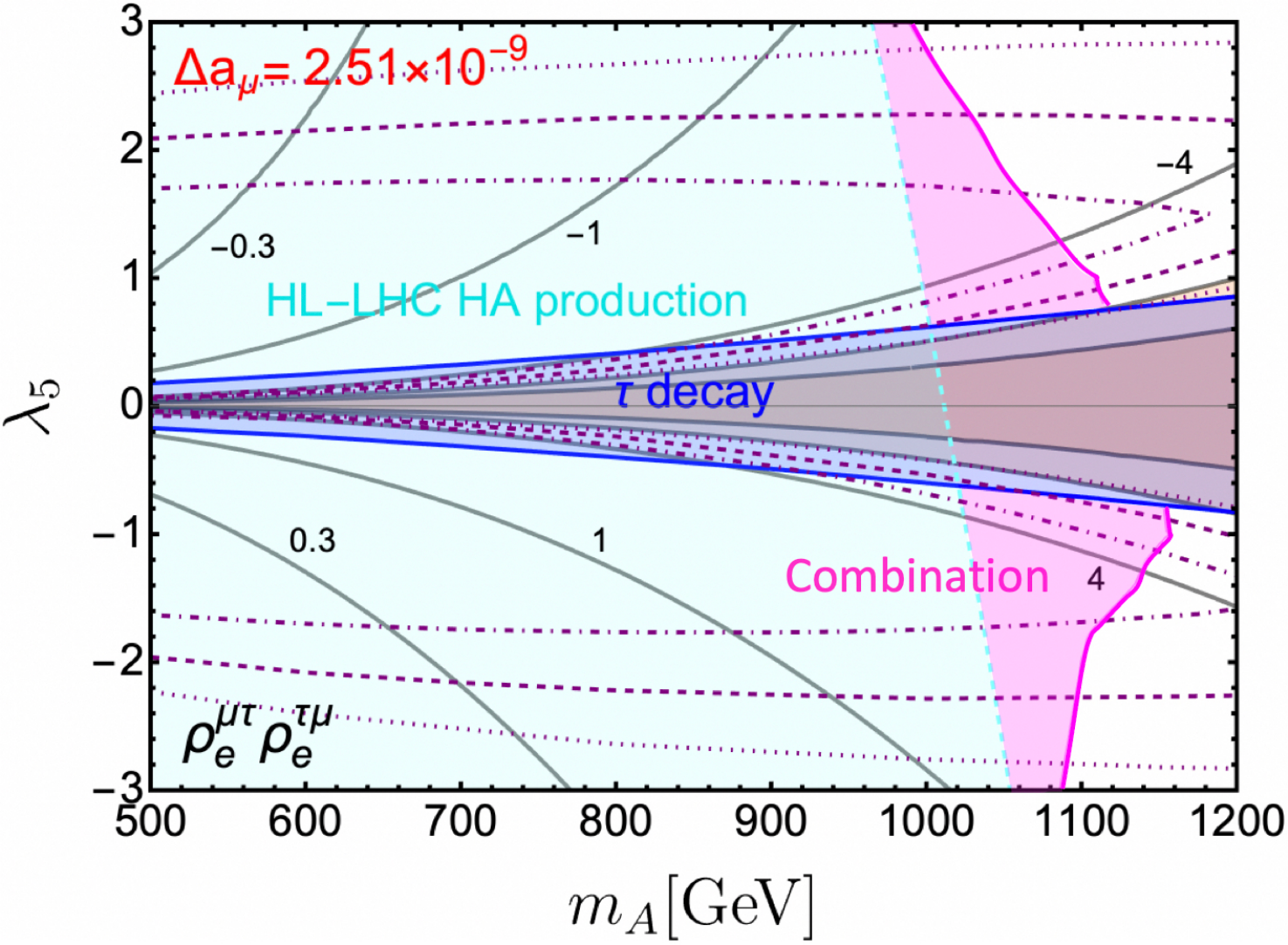}
     \vspace{-.55cm}
\caption{(Left) Fiducial  cross section of $\mu^\pm\mu^\pm\tau^\mp_h\tau^\mp_h$ via electroweak $HA$ production at a 14\,TeV $pp$ collider, as a function of $m_A$ in GeV. 
The horizontal gray lines correspond to the sensitivities with different luminosities.
(Right) The collider sensitivities are overlaid on Fig.\,\ref{fig:param}.
The cyan region can be tested with 3\,ab$^{-1}$  of the electroweak pair production, while the   magenta region can additionally be probed by including the single production channel.
The blue and orange regions and purple contours are the same as in Fig.\,\ref{fig:param}.
$\Delta a_\mu=2.51\times10^{-9}$ is fixed and $|\rho_e^{\mu\tau}|=|\rho_e^{\tau\mu}|$ is assumed in the figure.
        }
    \label{fig:14}
    \vspace{-.15cm}
\end{figure*}

Throughout our analysis, we use {\sc MadGraph}5\_a{\sc MC}@{\sc NLO} \cite{Alwall:2014hca} with NNPDF3.1luxQED \cite{Bertone:2017bme} to calculate the signal cross sections.
To account for the minimal kinematic cuts, $|p_T^\mu|$, $|p_T^\tau|$ $\ge 20$\,GeV, $|\eta_\mu|$, $|\eta_\tau| \le 2.7$, and $\Delta R \ge 0.1$ are imposed for all pairs of charged leptons.
We assume a hadronic $\tau$-tagging efficiency of $70\%$ \cite{ATLAS:2019uhp} and the hadronic $\tau$ decay branching ratio of about $65\%$ \cite{ParticleDataGroup:2020ssz}.
It is noted that an excellent $\tau$ charge reconstruction is reported in Ref.\,\cite{ATLAS:2015elo}.
Our signal of same sign $\mu^\pm\mu^\pm$ and $\tau^\mp\tau^\mp$ pairs of which $\mu^\pm\tau^\mp$ forms a resonance is very distinctive. Specifically, due to the resonance structure, the final-state leptons are very energetic.
Therefore we can safely assume the SM background (SMBG) to be negligible.\footnote{Background events come from $pp\to ZZ\to4\tau$. We confirmed that the contribution is smaller than $\mathcal{O}(10^{-5})$\,fb and can safely be neglected in the resonant regime. }
In this situation Poisson statistics is applicable and the sensitivity at $95\%$ confidence level is given when $\simeq3$ signal events are predicted, if no events are observed in the data.
The additional decay channels $H\to W^\pm H^\mp$ and $A\to HZ$ open in the case of large mass differences.\footnote{Note that the decay $A\to h Z$ is forbidden by the $\mathbb{Z}_4$ symmetry.}
However, such  large mass splittings are difficult for $\mathcal{O}(1)$\,TeV scalars.
We numerically included this dilution effect.

\begin{figure*}[t]
 \includegraphics[width=0.5\textwidth]{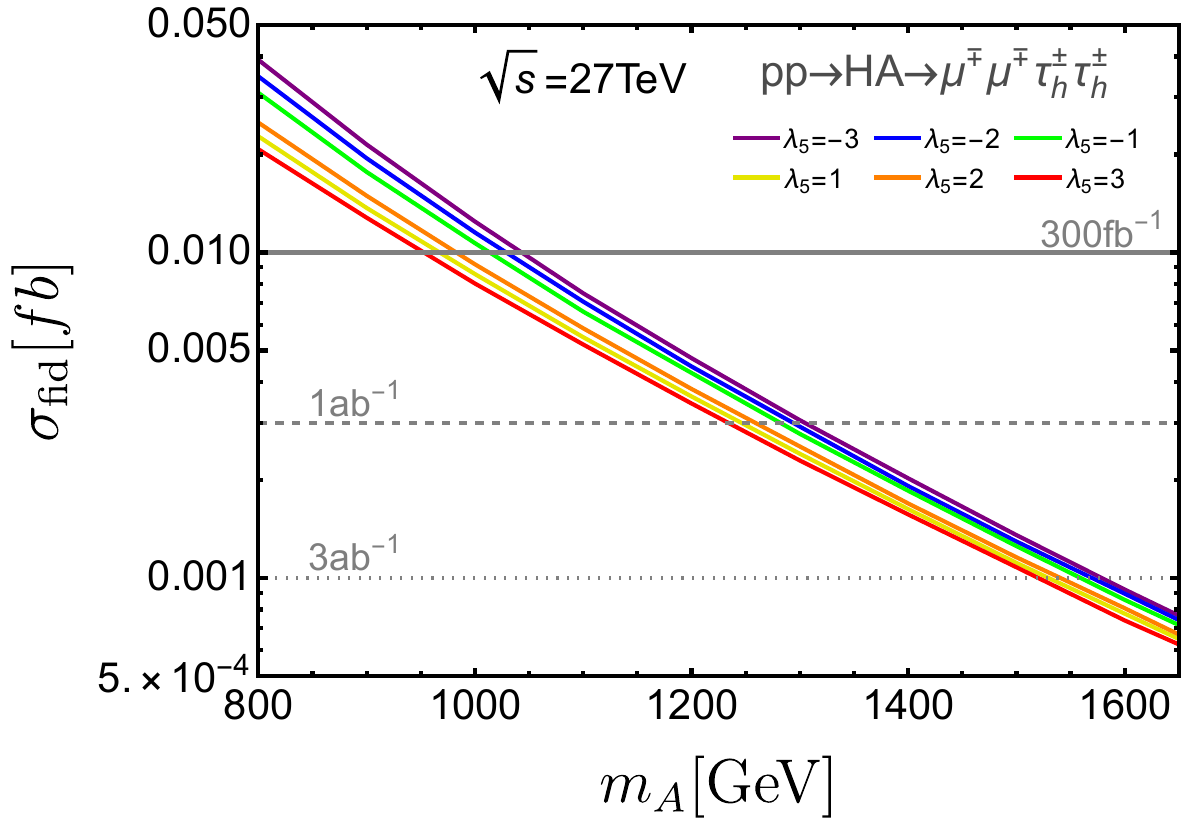}~
  \includegraphics[width=0.475\textwidth]{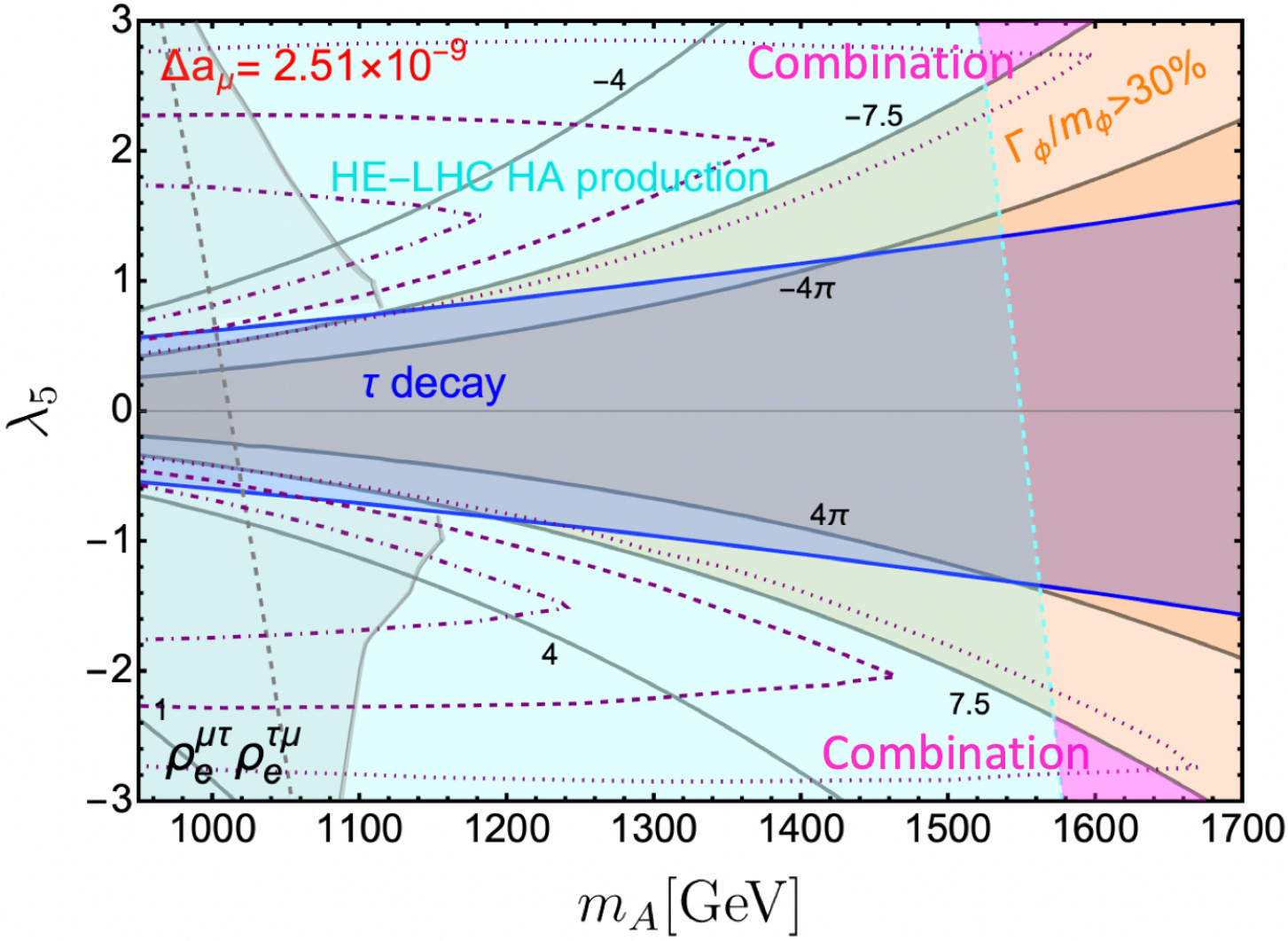}
     \vspace{-.4cm}
\caption{(Left)  Fiducial cross section of $\mu^\pm\mu^\pm\tau^\mp_h\tau^\mp_h$ via electroweak $HA$ production as a function of $m_A$ in GeV. The colored lines show the model prediction at a 27\,TeV $pp$ collider.
The horizontal gray lines correspond to the sensitivities with different luminosities.
(Right)  The cyan region shows the sensitivity of the HE-LHC based on $HA$ production. The magenta region can additionally be probed by including single production.
The gray region displays the sensitivity of the HL-LHC.
See also the caption in Figs.\,\ref{fig:param} and \ref{fig:14}.  
}
    \label{fig:27}
    \vspace{-.05cm}
\end{figure*}

Interpreting the DM candidate $S$ as the relic of a thermal freeze-out \cite{Asai:2022uix} requires a non-vanishing portal interaction between the Higgs doublets and $S$ in eq.\ \eqref{eq:potential},
\begin{align}
    V\supset \kappa(H_1^\dagger H_2)S^2+ {\rm h.c.}\,.
\end{align}
This coupling induces the invisible scalar decay $\phi\to SS$, where $\phi=H,A$. In turn this
 would suppress BR($\phi\to\mu\tau$) and hence reduce the $\mu^\pm\mu^\pm\tau^\mp_h\tau^\mp_h$ signal number.
However, BR($\phi\to\mu\tau$) is proportional to $m_\phi$ while BR($\phi\to SS$) is proportional to $m_\phi^{-1}$ because of the dimensionful coupling $\kappa v$ originating from the portal interaction.
As a result BR($\phi\to SS$) is suppressed in the heavy scalar regime and BR$(\phi\to\mu\tau)\simeq1$ holds well.
Therefore the $\mu^\pm\mu^\pm\tau^\mp\tau^\mp$ mode remains a viable and important probe of the model also in the presence of the DM candidate $S$, underlining the broad applicability of this search channel.
It is noted that the relic abundance requires $\kappa$ of $\mathcal{O}$(0.1) and hence its presence does not  have a significant impact on the phenomenology discussed in the present paper.

The colored contours in Fig.\,\ref{fig:14} (left) show the fiducial $\mu^\pm\mu^\pm\tau^\mp_h\tau^\mp_h$ cross sections based on the electroweak $HA$ production with $\sqrt{s}=14\,$TeV.
The horizontal lines correspond to the sensitivity for integrated luminosities of 139\,fb$^{-1}$, 1\,ab$^{-1}$, and 3\,ab$^{-1}$, and
correspond to cross sections of 0.02\,fb, 0.003\,fb and 0.001\,fb, respectively. For these cross sections $\simeq 3$ signal events are expected at the respective integrated luminosities so that an exclusion at 95\% confidence level can be obtained in the absence of signal.

Therefore, the HL-LHC data of 1\,ab$^{-1}$ and 3\,ab$^{-1}$ would 
be sensitive to mass scales of $m_A\simeq 800$\,GeV and $m_A\simeq 960$\,GeV.
It is worth mentioning that the pair production channel is powerful since once the neutral scalars are produced they dominantly decay into $\mu\tau$, as long as a sizable $\tau$-mass enhanced contribution to $a_\mu$ is postulated.
Nevertheless there is a mass gap between the sensitivity and the theoretical upper limit of Eq.\,(\ref{eq:thUL}).
The loss of sensitivity for larger $m_A$ mainly comes from two factors:
the contributing coupling constant is a weak gauge coupling which is independent of $\Delta a_\mu$ and the production cross section is suppressed by the heaviness of the pair-produced scalars.

One possible way to extend the LHC reach to our model is to include the single heavy scalar production channels corresponding to the middle and right diagrams of Fig.\,\ref{fig:dia}.
Especially for $\mathcal{O}(1)$\,TeV lepto-philic particles the inclusion of the photon initiated process is important \cite{Iguro:2020qbc}.
Again, following the procedure in Ref.\,\cite{Altmannshofer:2016brv} we assume the SMBG to be negligible.
There are two terms in the single scalar production amplitude to generate $\mu^\pm\mu^\pm\tau^\mp\tau^\mp$ events.
One is proportional to $(|\rho_e^{\mu\tau}|^2+|\rho_e^{\tau\mu}|^2)$ which vanishes in the $m_A=m_H$ limit.
On the other hand the term proportional to $\rho_e^{\mu\tau}\rho_e^{\tau\mu}$ does not disappear in this limit.
Since the $m_A^{-4}$ scaling in Eq.\,(\ref{eq:gm2}) requires a large product of the LFV Yukawa couplings and the first term is suppressed by the mass difference, the second contribution will be important for $\mathcal{O}(1)\,$TeV scalars.

In Fig.\,\ref{fig:14} (right) we overlay the collider sensitivity in the $m_A$ vs. $\lambda_5$ plane.
The cyan region shows the sensitivity of the pair production channel.
The sensitivity is asymmetric in $\lambda_5$, since $\lambda_5\ge0$ corresponds to $m_H\ge m_A$ and thus the production cross section will be smaller compared to $m_H\le m_A$.
The magenta regions can additionally be covered by including also the single scalar production. We note that there is also a non-resonant signal contribution which comes from $t$-channel $A/H$ exchange.
Since the lepton $p_T$ in this case is generally small and we are interested in the high $p_T$ region where the BG is negligible, this contribution is separated and subtracted to evaluate the sensitivity.
We find that the inclusion of the single production process can improve the experimental reach by 130 and 60\,GeV for $|\lambda_5|\simeq 1$ and 2, respectively, when the Yukawa couplings are fixed to explain the central value of the muon $g-2$.
This 130\,GeV gain approximately reduces to 90\,GeV (50\,GeV) when the size of the NP contribution in $g-2$ is reduced by $-1\,\sigma$ ($-2\,\sigma$).
This still leaves a gap between the experimental reach and theoretical upper limit.

In order to further boost the sensitivity to the large-mass scenario it is important to increase the center of mass energy from $\sqrt{s}=14$\,TeV to, for instance, 27\,TeV \cite{FCC:2018bvk}.
The colored lines in Fig.\,\ref{fig:27} (left) show the fiducial pair production cross section with $\sqrt{s}=27$\,TeV.
The horizontal lines correspond to the sensitivity for integrated luminosities of 300\,fb$^{-1}$, 1\,ab$^{-1}$, and 3\,ab$^{-1}$.
Thanks to the larger center of mass energy we see that 1250\,(1550)\,GeV can be covered with 1\,(3)\,ab$^{-1}$ of  data.
It is noted that the same kinematic cut introduced above has been applied for simplicity.
The sensitivity is shown in cyan in the right panel.

Again the reach of the $\mu^\pm\mu^\pm\tau^\mp\tau^\mp$ channel can be extended by including the single production process. The magenta regions in Fig.\,\ref{fig:27} (right) in the right corners can also be probed.
Compared to the sensitivity with $\sqrt{s}=14\,$TeV shown in gray on the left, the high energy (HE)-LHC is significantly more sensitive to the heavy scalar scenario. 
As a result, we observe that all the theoretically viable parameter region in Fig.\,\ref{fig:27} (right) can be covered.

Finally, a comment is in order concerning the impact of the assumptions on the Yukawa couplings $\rho_e^{\mu\tau}, \rho_e^{\tau\mu}$ entering our analysis.  As mentioned in Section \ref{sec:model} for given scalar masses $m_\phi$ their effect in $a_\mu$ is maximized for equal couplings $|\rho_e^{\mu\tau}|=|\rho_e^{\tau\mu}|$ and we hence restricted our attention to this limit. While moving away from the equal coupling scenario decreases $\Delta a_{\mu}$, the $HA$ pair-production cross section and the branching ratio into the $\mu^\pm\mu^\pm\tau_h^\mp\tau_h^\mp$ final state remain unaffected. As a consequence, for $|\rho_e^{\mu\tau}|\ne|\rho_e^{\tau\mu}|$ the $HA$ pair-production channel becomes more sensitive to the scalar solution to the $g-2$ anomaly in this model. The single-production cross section, as discussed above, decomposes into two contributions, the dominant of which is proportional to the same product $\rho_e^{\mu\tau}\rho_e^{\tau\mu}$ as $\Delta a_\mu$. The sensitivity of this channel is therefore independent on the relative size of $\rho_e^{\mu\tau}$ and $\rho_e^{\tau\mu}$, as long as $\Delta a_\mu$ is fixed. It will on the other hand decrease (increase) if eventually a smaller (larger) NP contribution to $g-2$ is required.

\section{Summary and Discussion}
\label{sec:summary}
The new FNAL experimental data for the muon $g-2$ is consistent with the previously measured value at the Brookhaven experiment, and the significance of the long-standing discrepancy now amounts to 4.2\,$\sigma$.
In this article we revisited the collider prospects of the $\mathbb{Z}_4$-based 2HDM which can explain the discrepancy using new one-loop contributions involving $\tau$ and neutral scalars.
A distinctive model prediction is the $\mu^\pm\mu^\pm\tau^\mp\tau^\mp$ signature at the LHC.
Since the viable parameter space of the model can not be fully probed at the LHC with the previously proposed electroweak scalar pair production, we investigated the impact of the single scalar production.
We have shown that the latter mode helps to extend the reach for the $\mathcal{O}(1)$\,TeV scalar solution of the muon $g-2$ discrepancy.
For instance, the HL-LHC with 3\,ab$^{-1}$ can test up to $m_A=1100\,$GeV. 
We also examined the search potential at the HE-LHC and showed that increasing the center of the mass energy is crucial to fully probe our scenario.
While the model can also explain the DM relic abundance, the reach of the proposed search does not depend on the DM interpretation. 

In this article we did not discuss $H^\pm\phi$ production. 
Due to a combinatorial factor the cross section is four times larger than the one  of $HA$ pair production \cite{Iguro:2019sly}.
The final state contains three charged leptons $2\tau+\mu$ or $2\mu+\tau$ and a neutrino at parton level, with their relative rates depending on the ratio of $|\rho_e^{\mu\tau}|^2$ and $|\rho_e^{\tau\mu}|^2$ \cite{Wang:2019ngf}.
Hence $H^\pm\phi$ production is expected to have better sensitivity to  $|\rho_e^{\tau\mu}|^2$.
Allowing for an imbalance in the couplings, $\rho_e^{\tau\mu} < \rho_e^{\mu\tau}$, while keeping the product $\rho_e^{\tau\mu} \rho_e^{\mu\tau}$ fixed, will thus decrease the LHC sensitivity to the $g-2$ solution in  $H^\pm\phi$ production. 
At the same time the required larger value of $\rho_e^{\mu\tau}$  also lowers the cut-off scale $\Lambda_\text{LP}$ of the theory. 
Note that the single scalar production cross section, being proportional to $(|\rho_e^{\mu\tau}|^2+|\rho_e^{\tau\mu}|^2)$, is enhanced in this case.
Thus, the signal discussed here is more universal.
Combining those various search channels to further enhance the sensitivity would be an interesting future direction.

Motivated by the muon $g-2$ discrepancy, we focused on the 2HDM with $\mu\tau$ LFV couplings. 
While the $e\mu$- and $e\tau$-philic scenarios lack such motivation, LFV particles can be a viable DM mediator and they predict a similar collider phenomenology.
Especially the $e\mu$ case is attractive in this respect, since the particle reconstruction is easier and the fiducial cross section via electroweak pair production is larger by a factor of five since the hadronic $\tau$ decay branching ratio and tagging efficiency do no longer reduce the signal rate.
This means that the relevant cross sections can be obtained from Fig.\,\ref{fig:14} and Fig.\,\ref{fig:27} by lifting the predictions by a factor of five.\\

\section*{Acknowledgements}
We thank Ulrich Nierste, Shohei Okawa and Yuji Omura for valuable comments and discussion.
We enjoy the support from the Deutsche Forschungsgemeinschaft (DFG, German Research Foundation) under grant 396021762-TRR\,257.
We also appreciate KIT for providing the necessary computational resources.
M.B. thanks the Mainz Institute for Theoretical Physics (MITP) of the Cluster of Excellence PRISMA$^+$ (Project ID 39083149) for its hospitality and support during the completion of this project.

\bibliography{ref}
\end{document}